\documentclass[twocolumn,aps,prd,epsf]{revtex4}
\usepackage{epsfig}
\usepackage{dcolumn}

\newcommand{\be}{\begin{equation}}
\newcommand{\ee}{\end{equation}}
\newcommand{\bea}{\begin{eqnarray}}
\newcommand{\eea}{\end{eqnarray}}

\begin{document}

\title{ Microscopic quark study of the $\eta$ and $\eta'$ masses }
\author{O. Lakhina}
\email{olakhina@kent.edu} \affiliation{Center for Nuclear Research,
Department of Physics, Kent State University, Kent OH 44242, USA}
\author{P. Bicudo}
\email{bicudo@ist.utl.pt}
\affiliation{CFTP, Dep. F\'{\i}sica, Instituto Superior T\'ecnico,
Av. Rovisco Pais, 1049-001 Lisboa, Portugal. }
\begin{abstract}
We show that it is necessary to go beyond the BCS (rainbow-ladder)
approximation  to split the $\eta$ and $\eta'$ masses from the $\pi$
and $K$ masses. We determine the self-consistent set of
one-quark-loop diagrams both for the Schwinger-Dyson quark mass gap
equation and for the Bethe-Salpeter quark-antiquark boundstate
equation. We identify the dominant diagrams, and we focus on the
boundstate equation. We detail the Bethe-Salpeter equation, adding
the dominant new diagram to the BCS kernel. The relevant numerical
techniques are also discussed. The ideal cases of one, two and three
light flavors, relevant to lattice QCD are also explored, together
with the case of realistic current quark masses.
\end{abstract}

\maketitle

\twocolumngrid

%
%
%
%
%
%
%
%
%
%
\section{Introduction}\label{intr}

%
\begin{figure}[b]
\includegraphics[width=1.05\columnwidth]{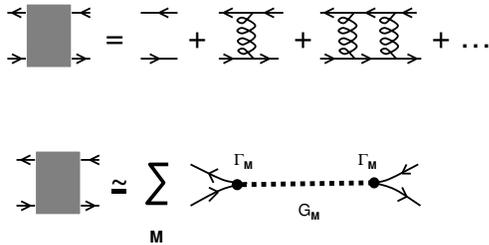}
\caption{ We show the ladder, minimal geometric series of diagrams
to include the mesons as Bethe-Salpeter quark-antiquark bound
states. This ladder is isopin invariant, i.e. in the chiral limit
the spectrum is identical for the $\pi$, the $K$, the $\eta$ and the
$\eta '$. Notice that, starting in this figure, for simplicity we
depict in the same way the bare and dressed gluon-quark-quark
vertices. }\label{Ladder}
\end{figure}

Many years ago Weinberg \cite{Weinberg:1975ui} noticed the
difficulty to account for the $\eta$ - $\eta'$ mass difference. He
baptized this difficulty as the U(1) problem.  Here we address the
U(1) problem in the microscopic perspective of quark models. Notice
that all effective models of QCD faced this problem. In their second
paper, Nambu and Jona-Lasinio \cite{Nambu:1961fr} include the
isospin-dependent pseudoscalar contact interactions in their model
to account for the $\eta$ - $\eta'$ mass difference, which are
reminiscent of scalar and pseudoscalar meson exchanges. In this
sense it is similar to the original $\sigma$ model of Gell-Mann and
L\'evy \cite{GellMann:1960np} , where the pseudoscalar exchange does
not include the isosinglet. A different determinant interaction was
invented by 't Hooft \cite{'t Hooft:1976fv}
 to split the
$\eta$ and $\eta'$ masses. With three flavors this is a tree-body
interaction. Recently a four-body interaction was further added by
Osipov, Hiller and Provid\^encia \cite{Osipov:2005tq} to stabilize
the vacuum. At a more fundamental level is Lattice QCD, where both
the quark and gluon fields are considered. However the Lattice QCD
simulations only reproduce the $\eta$ - $\eta'$ experimental mass
difference with the indirect technique
\cite{Duncan:1996ma,Kuramashi:1994aj} of the hairpin diagram
\cite{Witten:1979vv}, or with the extrapolation of the current quark
mass \cite{Aoki:2006xk}. we aim to understand how the $\eta$ -
$\eta'$ mass difference arises in a microscopic Quark Model .

%
\begin{figure}[t]
\includegraphics[width=1.05\columnwidth]{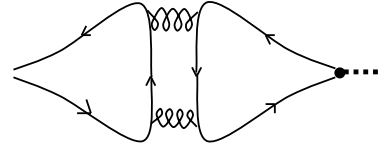}
\caption{
We show the minimal figure necessary to split the $\eta$ mass from the $\eta '$ mass.
This diagram, or any other diagram where the {\em in} fermion lines are not
continued to the {\em out} fermion lines, only contributes to $I=0$ mesons.
For other mesons it cancels.
}\label{MinimalKernelBBCS}
\end{figure}

In the Quark Model,
and in other related frameworks such as the Schwinger-Dyson
equations, the minimal computations of the meson masses are
performed in the {\em ladder} approximation. The ladder
geometric series is depicted in Fig. \ref{Ladder}.
In this case the
kernel (the set of diagrams that is iterated in the Bethe-Salpeter
equation) is a one-gluon exchange, or an effective quark-antiquark
interaction resulting from the integration of the gluon field.
In the limit of equal quark masses, the ladder approximation
does not separate the $\eta$  from the $\eta'$. Notice that any consistent
quark model should also include consistently chiral symmetry breaking.
Then, in the mass gap equation, the ladder approximation is
equivalent to the {\em rainbow}, or {\em BCS} approximation.
In the BCS approximation, the chiral symmetry is not a $SU(N_f)$
symmetry (where $N_f$ is the number of light fermions), but a
$U(N_f)$ global symmetry.

To address the $\eta$ - $\eta'$ mass difference problem one has to
go beyond the BCS approximation \cite{Bicudo:1998mc,Fischer:2007ze}.
It is necessary to add at least one quark loop in the mass gap
equation, including quark-antiquark annihilation and creation in the
quark-antiquark boundstate equation. The minimal extension to the
kernel of the boundstate equation is shown in Fig.
\ref{MinimalKernelBBCS}. Notice that this diagram includes two $AVV$
triangles, and thus includes the Adler-Bell-Jackiw anomaly
\cite{Adler:1969gk,Adler:1969er,Bell:1969ts}, in this case the
non-abelian anomaly, also related to the $U(1)$ problem \cite{'t
Hooft:1976up}. The anomaly is an ultraviolet effect and it is a
relevant effect. For instance for $\eta$, the anomalous
electromagnetic decay to $\gamma \gamma$ is of the same order as the
hadronic decays.

There has been an attempt to compute the diagram of Fig.\ref{Ladder}
in SDE approach and explore its effect on $\eta$ - $\eta'$ mass
difference \cite{von Smekal:1997dq}. However, certain assumptions
were made about the infrared behavior of the gluon propagator, which
have been shown later not to be true \cite{von Smekal:1997is}.

\begin{figure}[t]
\includegraphics[width=1.05\columnwidth]{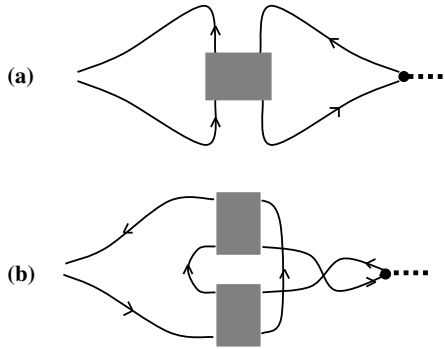}
\caption{ In non-perturbative QCD, the minimal diagram should be
dressed with all possible gluon exchange insertions. We show in (a)
the minimal one-ladder-figure necessary to split the $\eta$ mass
from the $\eta '$ mass. In (b) we show a two-ladder diagram,
equivalent to the coupling to the initial one-meson channel to the
two-meson channels. Both these diagrams only contribute to $I=0$
mesons. }\label{PossibleLadderDressingKernel}
\end{figure}

Moreover, in non-perturbative infrared QCD, the diagram of Fig.
\ref{MinimalKernelBBCS} is not really complete. The kernel should
include all diagrams of the same class \cite{Bicudo:1998mc}. In
particular the $\eta$ - $\eta'$ mass difference is a
non-perturbative problem and there is no reason to include only two
gluons, or two effective quark-antiquark interactions. Inasmuch as
the full geometric series included in the boundstate minimal
boundstate study, the diagram of Fig. \ref{MinimalKernelBBCS} should
be dressed with at least a full ladder series. In Fig.
\ref{PossibleLadderDressingKernel} (a) we include all the possible
number of gluon exchanges, and this is equivalent to include a full
meson-like ladder. This diagram includes all possible t-channel
exchanges of mesons.
%

\begin{figure}[t]
\includegraphics[width=1.05\columnwidth]{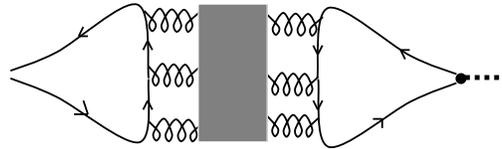}
\caption{
It is also possible to dress the minimal diagram with a pure-gluon ladder.
We show a different dressing, equivalent to couple the
pseudoscalar mesons to glueballs. Notice that with transverse gluons at
least three gluons are needed to constitute a pseudoscalar $J^{PC}=0^{-+}$
glueball.
This diagram also contributes to $I=0$ mesons, but it is suppressed by
the large three-gluon glueball masses.
}\label{GlueballsKernelBBCS}
\end{figure}

A different possible way to dress the diagram is depicted in Fig.
\ref{PossibleLadderDressingKernel} (b). In this case the diagrams
are resummed in two ladders. This is equivalent to couple the
original one meson to all possible two-meson channels. Notice that,
since a pseudoscalar cannot couple to two pseudoscalars, the
lightest coupled channel includes one pseudoscalar and one vector,
or one pseudoscalar and one scalar. So this channel is already quite
heavier than the original one meson channel of Fig.
\ref{PossibleLadderDressingKernel} (a).
In Fig. \ref{GlueballsKernelBBCS} we exchange all the possible
number of gluons, not between the quark lines, but between the gluon
lines, include a full glueball-like ladder. This diagram is
equivalent to couple the pseudoscalar mesons to glueballs. Notice
that with transverse gluons at least three gluons are needed to
constitute a pseudoscalar $J^{PC}=0^{-+}$ glueball. Both in lattice
QCD and in constituent quark models, including models where the
gluon mass is generated with a mass gap or Schwinger-Dyson equation,
the three gluon glueballs are quite heavy, much heavier than the
mesonic coupled channel of one pseudoscalar and one vector.
%
Thus we estimate that the leading non-perturbative diagram
contributing to the $\eta - \eta'$ mass difference is the t-channel
one meson exchange represented in Fig. \ref{PossibleLadderDressingKernel} (a).
In particular, because the pion is a light meson, the exchange of a
virtual pion should be relevant. This is consistent with the results
of the Nambu and Jona-Lasinio model and of the $\sigma$ model.

Importantly, chiral symmetry forces us to use a self-consistent set
of diagrams. Similarly to the sets of diagrams that preserve gauge
invariance when a photon is coupled, in the study of pseudoscalar
mesons a self-consistent set of diagrams is necessary.

Notice that different frameworks may be used to compute the necessary
diagrams. We may address the U(1) problem in the equal-time quark model
formalism, or we may use the related euclidian-time Schwinger-Dyson
formalism. Although the quark model is explicitly confining, and
able to reproduce the hadronic spectra up to high excitations, the
separation of the quark and antiquark propagators
\cite{Bicudo:1998mc}
would force us to use a much too large number of diagrams. Thus we will
work in the Schwinger-Dyson formalism, with Euclidian momenta integrations,
expecting that our results will also apply to the quark model.

In Section II we determine the necessary self-consistent set of
diagrams both for the Schwinger-Dyson quark mass gap equation and
for the Bethe-Salpeter quark-antiquark boundstate equation. Because
the full set of diagrams goes beyond the present state of the art
techniques, in Section III we identify the dominant diagrams. Notice
that the effect of meson exchange in the mass gap equation has
already been estimated, and thus we focus here on the boundstate
equation only. We also detail the Bethe Salpeter equation, adding
the dominant new $U(1)$ splitting diagram to the BCS kernel. In
Section IV we discuss the relevant numerical techniques and we show
the results for the $\eta$ and $\eta'$ masses. The ideal case of two
light flavors, relevant to lattice QCD is also explored, together
with the case of realistic current quark masses with three flavors.
In section V we conclude.

\section{A chirally self-consistent class of diagrams}\label{chiral}

%
\begin{figure}[t]
\includegraphics[width=1.05\columnwidth]{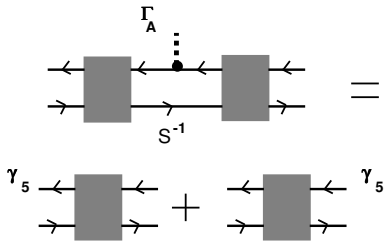}
\caption{
We show the ladder Ward Identity, crucial to verify that the Schwinger Dyson
mass gap equation is consistent with the Bethe Salpeter bound state equation.
}\label{WardIdentityLadder}
\end{figure}

We choose the Schwinger-Dyson equation (SDE) formalism to
investigate the effect of the diagrams that contribute differently
to the masses of isoscalars ($\eta$ or $\eta'$) and isovector
($\pi^0$). For recent reviews on the SDEs and their use in hadron
physics, see for example \cite{Holl:2006ni,Maris:2003vk}.

The SDEs are an infinite number of coupled integral equations;
a simultaneous, self-consistent solution of the complete set is
equivalent to a solution of the theory. In practice, the complete
solution of SDEs is not possible for QCD. Therefore one employs a
truncation scheme by solving only the equations important to the
problem under consideration and making assumptions for the solutions
of the other equations.

The simplest Schwinger-Dyson equation is the gap equation for the
quark propagator. It provides the relationship between the quark
propagator, the gluon propagator and the quark-gluon vertex. The
exact form of this equation is,
\begin{eqnarray}\label{SDE}
S(p)^{-1}&=&Z_2\left(i\gamma\cdot p+m_{bm}\right) +\\
&&Z_1\int\frac{d^4 q}{(2\pi)^4}g^2 D_{\mu\nu}(q)
\frac{\lambda^a}{2}\gamma_{\mu}S(p+q)\Gamma^a_{\nu}(q,p),\nonumber
\end{eqnarray}
where $S(p)$ is the flavor dependent fully dressed quark propagator,
which has the form: \be S(p)=\frac{1}{i\hspace{-.08cm}\not{
p}A\left(p^2\right)+B\left(p^2\right)}, \label{quarkProp} \ee
$D_{\mu\nu}(p-q)$ is the gluon propagator, and $\Gamma^a_{\nu}(q,p)$
is the quark-gluon vertex, $a$ denotes the flavor of the quark. The
propagators and the vertex in this equation are dressed and
renormalized. $Z_1$ and $Z_2$ are the renormalization constants of
the quark-gluon vertex and the quark wave function, and $m_{bm}$ is
the bare quark mass.

%
\begin{figure}[t]
\includegraphics[width=1.05\columnwidth]{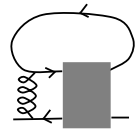}
\caption{
Contribution of the meson exchange kernel to the mass gap
Schwinger-Dyson equation
}\label{LadderMassGap}
\end{figure}

The Bethe-Salpeter equation is used to study two-particle bound
states (mesons). By solving the Bethe-Salpeter equation one obtains the
Bethe-Salpeter amplitude which, after proper normalization,
completely describes the meson as the bound state of a quark and an
antiquark. The Bethe-Salpeter amplitude corresponds to the amputated
one-particle-irreducible quark-meson vertex, and its Lorentz
structure depends on the quantum numbers of the meson of interest.
The exact Bethe-Salpeter equation for the meson can be written as,
\bea
\label{BSEMeson}
\Gamma_M(p_1,p_2)&=&\\
\int\frac{d^4
q}{(2\pi)^4}&&\!\!\!\!\!\!\!\!\!K\left(q\right)S(p_1+q)
\Gamma_M\left(p_1+q,p_2+q\right)S(p_2+q),\nonumber
\eea
where
$\Gamma_M(p;P)$ is the Bethe-Salpeter amplitude for the meson $M$.
The Bethe-Salpeter amplitude depends on the quark momenta $p_1$ and
$p_2$. If $m$ is the bound state mass then equation (\ref{BSEMeson})
is only valid for $P^2=(p_1-p_2)^2=-m^2$, where $P$ is the total
momentum of the meson. The kernel $K(q)$ is the irreducible
quark-antiquark scattering kernel.

The truncated mass gap SDE and the BSE have to be consistent with
each other to preserve the chiral symmetry of QCD. This
self-consistency is crucial for the pion to be a massless boundstate
in the chiral limit. It has been shown that if one inserts the Bethe
Salpeter vertex $\Gamma$ in all possible dressed quark propagators
$S(p)$ present in the SDE for the quark self-energy, one must
recover the kernel of the BSE. Inversely, if one removes the
Bethe-Salpeter vertex from the BSE, replacing $S \Gamma S$ by the
propagator $S$, one must recover the self-energy of the SDE. This
chiral self-consistent relation has already been applied to
self-energies and to Bethe Salpeter kernels containing the infinite
ladder series of diagrams
\cite{Bicudo:1998mc,Bicudo:2003fp,Llanes-Estrada:2003ha}.

Let us first show that using the rainbow-ladder truncation (BCS
approximation) one obtains a self-consistent set of equations for
quark propagator, bound states and vertices. After that, an analogous
set of equations will be derived for a different truncation, which
includes the diagrams necessary for generating $\eta$-$\eta'$ mass.

The rainbow truncation of Schwinger-Dyson equation for the quark
propagator includes the replacement of the dressed quark-gluon
vertex by the bare one,
\be \label{rainbow}
Z_1g^2D_{\mu\nu}(q)\Gamma^a_{\nu}(q,p)\,\rightarrow\,\frac{G(q^2)}{q^2}
T_{\mu\nu}(q)\frac{\lambda^a}{2}\gamma_{\nu}, \ee
where $G(q^2)$ is
the effective running coupling, $q$ is the gluon momentum, and
$T_{\mu\nu}(q)=\delta_{\mu\nu}-q_{\mu}q_{\nu}/q^2$ is the transverse
operator in Landau gauge. The quantity $T_{\mu\nu}(q)/q^2$ is the
free gluon propagator.

In the rainbow truncation the equation (\ref{SDE}) becomes,
\begin{equation}
S(p)^{-1}=S_0^{-1}(p)+\frac{4}{3}\int {d^4 q \over (2 \pi)^4}
\frac{T_{\mu\nu}(q)}{q^2}G(q^2) \gamma_{\mu} S(p+q) \gamma_{\nu}  \
, \label{SD}
\end{equation}
where $S_0(p)=\left(i{\not{p}}+m\right)^{-1}$ is the bare quark
propagator, $m$ is the current quark mass.

%
\begin{figure}[t]
\includegraphics[width=1.05\columnwidth]{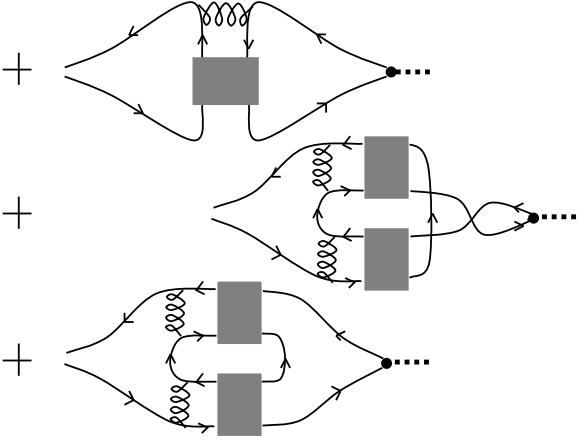}
\caption{
Contribution of the meson exchange kernel to the boundstate
Bethe-Salpeter equation
}\label{ThreeLadderandOGEKernelBBCS}
\end{figure}

The equation (\ref{SD}) increases the gap between the quark and
antiquark dispersion relations (which is almost zero for bare
propagators) and generates the dynamical mass $M=\sqrt{B^2(p^2) /
A^2(p^2)}$. Because of the color structure of the eqn. (\ref{SD}),
the tadpole does not contribute. Multiplying both sides of eqn.
(\ref{SD}) with $\gamma_5$ and summing leads to,
\begin{eqnarray}
&S&(p_1)^{-1}\gamma_5 + \gamma_5 S(p_2)^{-1} = S_0(p_1)^{-1}\gamma_5
+ \gamma_5 S_0(p_2)^{-1}\nonumber
\\
&-&\!\!\!\!\! \int\!\!\!{d^4 q \over (2 \pi)^4}
\frac{T_{\mu\nu}(q)}{q^2}G(q^2) \gamma_{\mu} \left[ S(p_1+q)\gamma_5
+ \gamma_5 S(p_2+q) \right] \gamma_{\nu},\nonumber
\end{eqnarray}
which is the Bethe Salpeter equation for the vertex,
\bea\label{BSEAxial}
\Gamma_{\hspace{-.08cm}A}(p_1,p_2)&=&\gamma_A(p_1,p_2)+\\
\int\frac{d^4
q}{(2\pi)^4}&&\!\!\!\!\!\!\!\!\!K\left(q\right)S(p_1+q)
\Gamma_{\hspace{-.08cm}A}\left(p_1+q,p_2+q\right)S(p_2+q),\nonumber\eea
if the bare $\gamma_{\hspace{-.08cm}A}$ and dressed
$\Gamma_{\hspace{-.08cm}A}$ vertices are respectively defined with
the same Axial Ward Identity for an isovector,
\begin{eqnarray}
\Gamma_{\hspace{-.08cm}A}(p_1,p_2) &=& S^{-1}(p_1) \gamma_5 +
\gamma_5 S^{-1}(p_2) \ , \label{axialWIdressed}\\
\gamma_{A}(p_1,p_2)&=&  S^{-1}_0(p_1) \gamma_5 + \gamma_5
S^{-1}_0(p_2) \ , \label{axialWIbare}
\end{eqnarray}
and the kernel of the BSE is approximated by the one-gluon exchange
(ladder truncation, or BCS approximation),
\be
\label{ladder}
K(q)\,\rightarrow\,-\frac{G(q^2)}{q^2}
T_{\mu\nu}(q)\frac{\lambda^c}{2}\gamma_{\mu}\otimes
\frac{\lambda^c}{2}\gamma_{\nu}.
\ee

The fact that Axial Ward identity is consistent with the ladder
approximation for the bound state and the rainbow approximation for
the quark self energy equation ensures that this truncation respects
the chiral symmetry of QCD. Both approximations are equivalent to
the planar diagram expansion which is characteristic of the Quark
Model.

\par
The bare vertex $\gamma_{A}$ can be computed from the bare quark
propagator using (\ref{axialWIbare}),
\begin{equation}
{\gamma_{A}}(P) =  (i\hspace{-.08cm}\not P  + 2 m ) \gamma_5 \ , \ \
P=p_1-p_2 \ , \label{bare}
\end{equation}
where $m$ is the current quark mass. $\gamma_{A}$ is the particular
part of the Bethe Salpeter equation for the vertex (\ref{BSEAxial}),
and it vanishes when the current quark mass $m$ is small (chiral
limit) and at the same time the total momentum $P^\mu$ of the vertex
is small.

%
\begin{figure}[t]
\includegraphics[width=1.05\columnwidth]{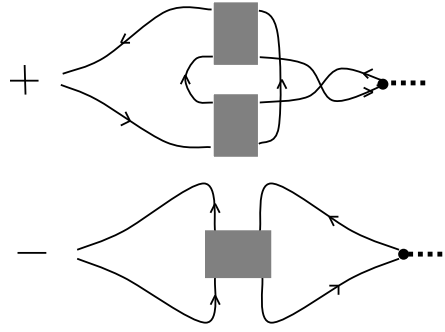}
\caption{ Non-perturbative part of the kernel contributing only to
flavor singlet $\eta'$ mass. }\label{twonewfinalkernels}
\end{figure}

\par
At this point it is important to clarify that in general, as it
follows from (\ref{bare}), the bare vertex $\gamma_A$ is a
combination of the axial vector vertex $\gamma^{\mu}\gamma_5$ and
the pseudoscalar vertex $\gamma_5$. In the chiral limit $\gamma_A$
becomes pure axial vector vertex contracted with momentum, and in
the limit of vanishing momentum $P_{\mu}$ it becomes pure
pseudoscalar vertex multiplied by the current quark mass. For
simplicity (and because they are defined with the Axial Ward
identity (\ref{axialWIdressed}) and (\ref{axialWIbare})), in the
rest of this paper $\gamma_{A}$ will be called the bare axial vertex
and $\Gamma_{\hspace{-.08cm}A}$ will be called the dressed axial
vertex, although they possess a more general Dirac structure.

The dressed vertex $\Gamma_{\hspace{-.08cm}A}$ can be computed from
the dressed quark propagator (\ref{quarkProp}) using
(\ref{axialWIdressed}),
\begin{equation}
\Gamma_{\hspace{-.08cm}A}(p_1,p_2) \hspace{-.08cm} = \left[ i A(p_1)
\hspace{-.1cm} \not p_1 \hspace{-.05cm} - \hspace{-.05cm} i A(p_2)
\hspace{-.1cm} \not p_2 \hspace{-.05cm} + \hspace{-.05cm} B(p_1)
\hspace{-.05cm} + \hspace{-.05cm} B(p_2) \right]\gamma_5 .
\hspace{-.3cm}
\end{equation}

$\Gamma_{\hspace{-.08cm}A}$ is finite providing spontaneous chiral
symmetry breaking occurs in eqn. (\ref{SD}) and a dynamical mass of
the quark is generated. For instance, if the total momentum
$P=p_1-p_2$  of the vertex vanishes, the vertex
$\Gamma_{\hspace{-.08cm}A}$ is simply identical to $2 B(p)
\gamma_5$, where $B(p)$ is a finite solution of the mass gap
equation. This vertex describes the coupling of the pion to two
quarks, and eqn. (\ref{BSEAxial}) shows that in the chiral limit we
have a Goldstone boson, the pion.

\par
For simplicity the flavor is not yet included. Flavor will only
be explicitly included at the end of subsection VII. The isoscalar
axial Ward identity must include the Axial anomaly, which is crucial to the $U(1)$
problem. Nevertheless the pion is an isovector, and in the coupling of a pion
we do not need to concern with the Axial anomaly.

%
\begin{figure}[t]
\includegraphics[width=1.05\columnwidth]{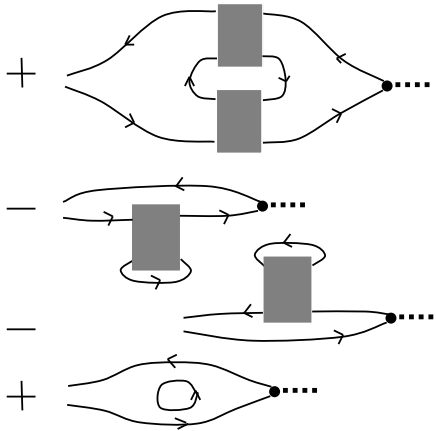}
\caption{ Non-perturbative part of the kernel shifting all the
pseudoscalar masses equally. }\label{fournewfinalkernels}
\end{figure}

\par
We now derive a powerful Ward identity for the ladder which involve
the axial vertices and the ladder. This relation is depicted in Fig.
\ref{WardIdentityLadder}, it constitutes an extension of the Ward
Identity for the propagators in eqs. (\ref{axialWIdressed}) and
(\ref{axialWIbare}). This identity is derived if we expand
\cite{Bicudo:1998mc,Bicudo:2003fp,Llanes-Estrada:2003ha} the ladders
and substitute the vertex in the left hand side. Then all terms with
an intermediate $\gamma_5$ include the anticommutator
$\{\gamma_5,\gamma_{_V}\}$ and this cancels because {\em the
interaction is chiral invariant} and {\em the kernel is local}. Only
the right hand side survives.

We are now ready to derive the mass gap and the boundstate equations
beyond rainbow-ladder truncation, to include the diagrams that are
able to generate $\eta$-$\eta'$ mass difference. Starting from the
diagram in Fig. \ref{PossibleLadderDressingKernel} (a) (which is the
minimal diagram that involves one $t$-channel meson exchange), and
substituting the Bethe-Salpeter vertex by a propagator, we arrive at
the contribution for the quark self-energy in Fig.
\ref{LadderMassGap}. Notice that, to remove any disconnected
diagrams, the first diagram must include at least one gluon
exchange, thus, in Fig. \ref{LadderMassGap}, the geometric series
starts with one gluon exchange.

Then, inserting the Bethe-Salpeter vertex in all possible
propagators of the quark self-energy, we get the kernel diagrams in
Fig. \ref{ThreeLadderandOGEKernelBBCS} for the boundstate equations.
The first kernel diagram is obtained inserting the vertex in the
only propagator of the self-energy exterior to the ladder. The
second kernel diagram is obtained inserting the vertex in the same
quark line, but inside the ladder of the self-energy. The third
kernel diagram is obtained inserting the vertex in the quark line
linking to the external legs of the self-energy. We can also express
these diagrams in terms of ladders only, and we finally get the
diagrams in Figs. \ref{twonewfinalkernels} and
\ref{fournewfinalkernels}. In Fig. \ref{twonewfinalkernels} we show
the diagrams contributing to the $U(1)_A$ mass splitting and in Fig.
\ref{fournewfinalkernels} we show the diagrams contributing to all
mesons.

\section{A calculable Bethe Salpeter Kernel}\label{kernel}

We now choose the best framework to compute the new beyond BCS
diagrams. Notice that the diagrams in Figs. \ref{LadderMassGap},
\ref{twonewfinalkernels} and \ref{fournewfinalkernels} all include
full ladders, and an integral in one of the relative variables of
the full ladder. Technically this remains a problem since it goes
beyond the present state of the art of quark models.

%
\begin{figure}[t]
\includegraphics[angle=90,width=1.0\columnwidth]{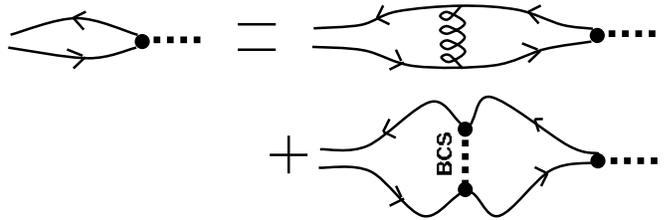}
\caption{ The homogeneous Bethe-Salpeter equation we solve here. The
kernel includes the BCS level One Gluon Exchange, the anomalous
annihilation into two gluons, and the non-perturbative One Pion
Exchange. }\label{finalBCSkernel}
\end{figure}

Two different classes of quark models, consistent with chiral
symmetry, have been investigated and applied to different problems.
The equal time quark models are confining, thus they are adequate
for the computation of the full ladder, including the full spectrum
of mesons. However they are not Lorentz invariant and thus are
inadequate for the boost of the ladder, occurring in the momentum
space integral in the Feynman diagrams. And the Lorentz
non-invariance also splits the time-like from the space-like
components and observables of the model. On the other hand, the
euclidian space quark models are convenient for full integrations of
internal momenta in Feynman diagrams, but are not confining and thus
are inadequate for the computation of the full ladder. However, the
full ladders can be replaced by the lowest boundstate contribution,
i. e. approximately assumed to be similar to the pseudoscalar pole
contribution. In this case the euclidian models, essentially
adequate to address the lowest energy phenomena of hadronic physics
(including the pseudoscalar ground states to the vector ground
states and pseudoscalar first excitations), can be applied to the
$U(1)_A$ breaking of the pseudoscalar spectrum.

The dominant part of the non-perturbative dressing of the kernel of
the Bethe Salpeter equation with hadron ladders is the One Meson
Exchange in the t-channel. Notice that the corresponding diagram
occurs with two minus signs when compared with the One Gluon
Exchange. A first minus sign is necessary to cancel all the
disconnected diagrams. A second minus sign is due to the fermion
loop.

\begin{figure}[t!]
\includegraphics[angle=0,width=1.\columnwidth]{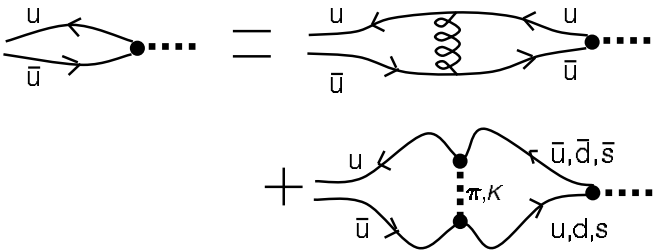}
\caption{An example of BSE equation needed to solve for the mass of
the bound state $u\bar{u}$. In the annihilation diagram quark
propagators of all flavors contribute. That makes it necessary to
include kaon exchange in this diagram if the strange flavor is taken
into account.}\label{uu-bar}
\end{figure}

Now, t-channel exchange is dominated by the pseudoscalar meson
exchange, since the pseudoscalar mesons are the lightest ones. Thus,
in this first attempt of the $\eta'$ mass study beyond BCS, we will
approximately saturate the meson exchange with only the ground state
pseudoscalar nonet. And, because we are mostly interested in the
$U(1)_A$ breaking, we will also neglect all terms that contribute
equally to all mesons, providing only a constant overall mass shift
of the different spectra. Thus we will not consider the contribution
to the mass gap equation of Fig. \ref{LadderMassGap} and the
contribution to the boundstate kernel in Fig
\ref{fournewfinalkernels} (it has also been estimated by one of us
\cite{Bicudo:1998mc} that these effects are relatively small). In
any case, these two effects cancel in the chiral limit, since our
self-consistent approach complies with the Goldstone theorem.

The minimal kernel necessary to estimate the different contributions
to the $\eta$ and $\eta'$ masses is presented in Fig.
\ref{finalBCSkernel}. The first diagram on the right-hand-side of
Fig. \ref{finalBCSkernel} is the BCS level One Gluon Exchange. This
diagram contributes to all mesons. The second diagram on the
right-hand-side includes the pion exchange in t-channel between
quark and antiquark. It only contributes to the isosinglet mesons,
and therefore generates the mass difference between the $\pi_0$, the
$\eta$ and the $\eta'$.

The diagrammatic form of Bethe-Salpeter equation depicted in Fig.
\ref{finalBCSkernel} can be written as,
\begin{eqnarray}\label{BSE}
\Gamma^{aa}_{\eta}(p;P)&=&\\
\int\frac{d^4
q}{(2\pi)^4}&\!\!\!K\!\!\!&\left(p,q;P\right)S^a(q+\eta P)
\Gamma^{aa}_{\eta}\left(q;P\right)S^a(q-\beta P)\nonumber\\
+\sum_{M,b}\int&\!\!\!\!\!\!\frac{d^4
q}{(2\pi)^4}\!\!\!\!\!\!&C^{ab}_M \left(p,q;P\right)S^b(q+\eta
P)\Gamma^{aa}_{\eta}\left(q;P\right)S^b(q-\beta P),\nonumber
\end{eqnarray}
where \bea C^{ab}_M\left(p,q;P\right)=\Gamma^{ab}_M(p,q,P) G_M(p-q)
\Gamma^{ba}_M(p,q,P), \eea $\Gamma^{ab}_M$ is the Bethe-Salpeter
amplitude and $G_M$ is the propagator of the meson being exchanged,
$a$ and $b$ denote quark flavors.

\begin{figure}[t!]
\includegraphics[angle=270,width=1.\columnwidth]{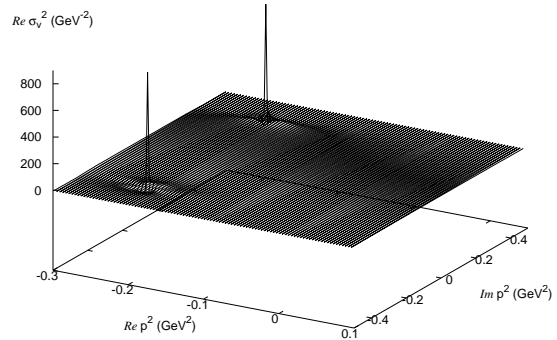}
\caption{Singularities of the solution of the gap equation for ${\it
u}$ quark. The real part of the scalar function $\sigma_v(p^2)$ is
presented, which is defined in (\ref{sigma}).}\label{UProp}
\end{figure}

The effective running coupling in the kernel can be modeled to
account for the effects of the truncation. We used the Maris-Tandy
model
\cite{Maris:1999nt,Maris:1997tm}
in our calculations,
\bea
\frac{G(k^2)}{k^2}&=&\frac{4\pi^2 D}{w^6}k^2 e^{-k^2/w^2} +\\
&&\frac{4\pi^2\gamma_m F(k^2)}
{\frac{1}{2}\ln\left[\tau+\left(1+k^2/\Lambda^2_{QCD}\right)^2\right]},
\eea
where $F(k^2)=\left(1-exp{\frac{-k^2}{4m_t^2}}\right)/k^2$,
$\gamma_m=12/\left(33-2N_f\right)$ an where the parameters are
$m_t=0.5$ GeV, $N_f=4$, $\Lambda_{QCD}=0.234$ GeV, $\tau=e^2-1$, $D=0.93$ GeV$^2$
and $w=0.4$ GeV. This model has been shown to work well for the
description of the light meson properties (see for example
\cite{Maris:2000sk,Maris:1997hd}).
Therefore we expect it to give
sensible results for the masses of $\eta$ and $\eta'$.

\begin{figure}[t!]
\includegraphics[angle=0,width=1.\columnwidth]{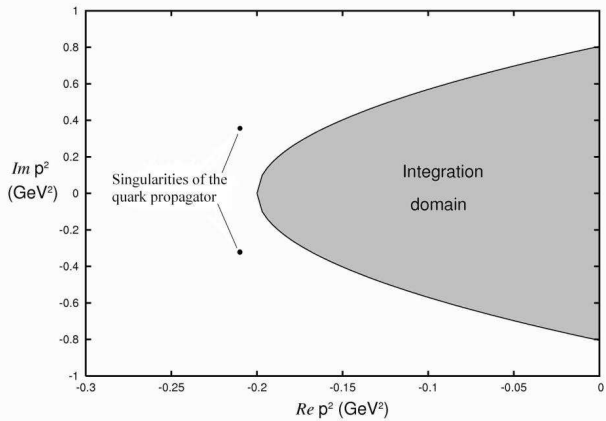}
\caption{Location of the singularities of the quark propagator with
respect to our largest integration domain of Bethe-Salpeter
equations.}\label{IntDom}
\end{figure}

In general, for $N_f$ quark flavors there will be $N_f$ coupled
integral Bethe-Salpeter equations, which have to be solved together.
An example of such an equation for the $u\bar{u}$ bound state with
all possible pseudoscalar meson exchanges is depicted in Fig.
\ref{uu-bar}.

Notice that one of the features of Schwinger-Dyson equation approach
is the fact that the solution for the quark propagator has
singularities. An example of the solution for the function
$\sigma_v(p^2)$ for the ${\it u}$ quark, \be
\sigma_v(p^2)=\frac{A(p^2)}{p^2A^2(p^2)+B^2(p^2)}.\label{sigma} \ee
is presented in Fig. \ref{UProp}. In general, these singularities
can pose a serious technical problem for the numerical calculations
if present in the integration domain of the integral Bethe-Salpeter
equations. However, we have checked that in our case these
singularities are outside from the largest integration domain that
we have. Our largest integration domain and the positions of the
poles are shown in Fig. \ref{IntDom}.

Thus the masses of the $\eta$ and $\eta'$ mesons are computable in the
SDE and BSE formalism, extending the Klabucar and Kekez parametrization
of the $U_a(1)$ breaking
\cite{Klabucar:1997zi}.

\section{Results}\label{results}

We solve the Bethe-Salpeter equation (\ref{BSE}) for the pseudoscalar
mesons. This equation takes into account the additional
``annihilation" diagram which contributes to the isoscalar but not
to the isovector mesons.

This additional diagram is depicted in Fig.
\ref{PossibleLadderDressingKernel}a, it consists of the annihilation
of our $q\bar{q}$ pair into an infinite gluon ladder. If calculated
perturbatively, this diagram does not contribute to the isoscalar
meson mass in the chiral limit. However, it gives rise to the
$\eta'$ meson mass if treated nonperturbatively. We propose to
represent the infinite gluon ladder in the ``annihilation" diagram
as a t-channel meson exchange. In general, mesons in a t-channel can
have any quantum numbers. However, the pion and the other
members of the pseudoscalar octet, being the lightest mesons,
give the dominant contribution.

\begin{figure}[t]
\includegraphics[angle=270,width=1.0\columnwidth]{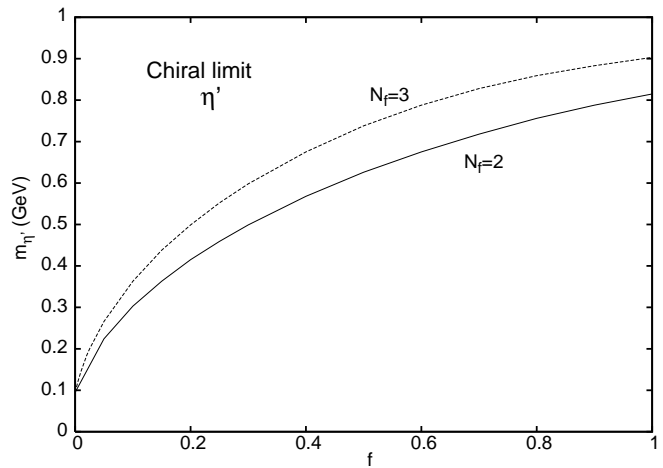}
\caption{Dependence of the mass of $\eta'$ meson on the "weight" of
the additional diagram in the chiral limit for two and three quark
flavors.} \label{EtaChiralLimit}
\end{figure}

This $U_A(1)$ breaking term
is added to the ladder Bethe-Salpeter kernel in the
Bethe-Salpeter equation (\ref{BSE}), as in Fig. \ref{finalBCSkernel}.
The Bethe-Salpeter amplitude, needed for the pseudoscalar-exchange kernel,
is previously calculated by solving Bethe-Salpeter equation in the ladder
truncation (\ref{ladder}). The quark propagator, which is needed for the
solution of the equation (\ref{BSE}) is also obtained by solving the
ladder Schwinger-Dyson equation (\ref{SDE}) numerically. This ensures
that the pion and kaon non-isoscalar masses are unaffected.
Having thus defined both the Bethe-Salpeter kernel and the quark propagator,
we solve numerically the Bethe-Salpeter equation by the power method.

First, we investigate the isoscalar meson mass in the chiral limit.
The chiral limit is the limit where the current quark masses vanish.
In this case the "annihilation" diagram is simply proportional to the
number of flavors $N_f$. From the extent of the $U_A(1)$ breaking it
is expected, that when the chiral limit is taken for all three quark
flavors, the $\pi_0$ and $\eta$ should become massless, and the $\eta'$
should remain massive. We are able to reproduce this behavior with our
phenomenological model: the "annihilation" diagram does not contribute to
the $\pi_0$ and $\eta$ in the chiral limit leaving these mesons massless,
but changes the mass of the flavor scalar $\eta'$ (Fig. \ref{EtaChiralLimit}).

\begin{figure}[t!]
\includegraphics[angle=270,width=1.\columnwidth]{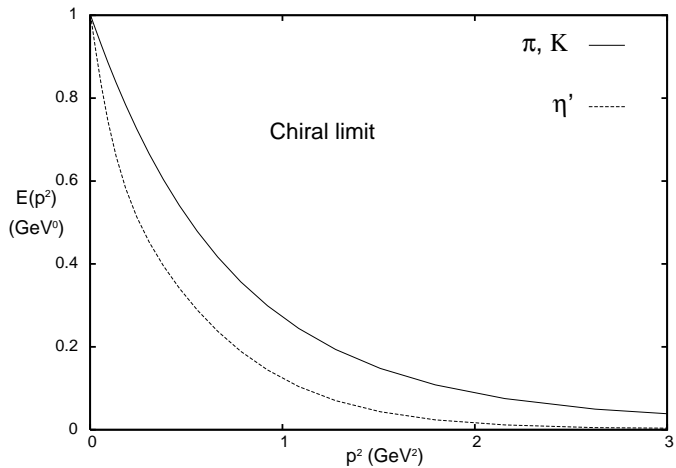}
\caption{The $E(p^2)$ Bethe-Salpeter amplitudes, solutions of
Bethe-Salpeter equation, for the $\pi$, the $K$ and the $\eta'$.}
\label{BSA_E}
\end{figure}

The Bethe Salpeter amplitude of pseudoscalar mesons can be
decomposed into four Dirac structures $E, \ F, \ G, \ H$. By keeping
only the $E$ amplitude  (separable in the chiral limit and dominant
for realistic masses) we calculate the $\eta'$ meson mass in the
chiral limit with the dependence on the "weight" of the additional
diagram (phenomenological factor $f$ which multiplies the additional
$U_A(1)$ breaking diagram in the BSE). The result for 2 and 3 quark
flavors is presented in Fig. \ref{EtaChiralLimit}. In the two-flavor
case there is only one $\eta$, in Fig. \ref{EtaChiralLimit} it is
also denoted $\eta'$. One can see that
our calculations give a very reasonable estimate of the mass of
$\eta'$ in the chiral limit. The solutions of the Bethe-Salpeter
equation for $\pi$, $K$ and $\eta'$ (in the chiral limit) are
presented in Fig. \ref{BSA_E}. Notice that in the chiral limit the
pion and kaon amplitudes are identical.

Next, in our calculation, we employ broken $SU(3)_f$ symmetry with {\it u}
and {\it d} quarks of equal finite mass, and realistically heavier
{\it s} quark. That means that in the "annihilation" diagram we have
to include both kaon and pion propagators. We examine the effect that
this diagram has on the physical mesons.
It has been shown before \cite{Maris:1997hd} that the SDE and BSE approach,
in the rainbow-ladder truncation, works reasonably well for the
properties of the light isovector mesons, and we reproduce these
results obtaining for the pion $m_{\pi}=0.139$ GeV and
$f_{\pi}=0.131$ GeV.

\begin{figure}[t!]
\includegraphics[angle=270,width=1.\columnwidth]{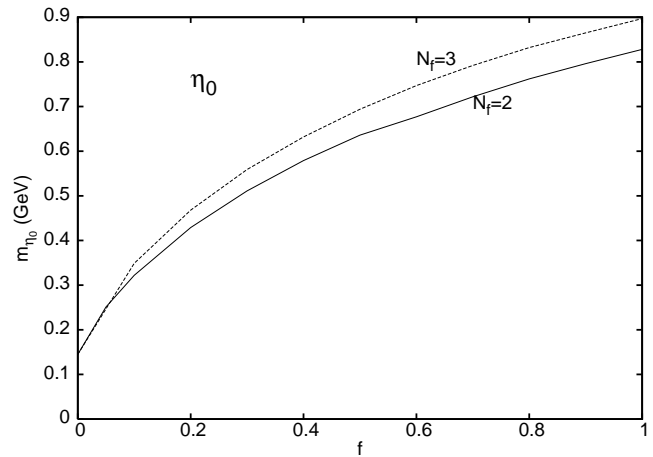}
\caption{Dependence of the flavor scalar $\eta_0$ mass on the "weight" factor {\it
f} for two and three quark flavors.}\label{eta0}
\end{figure}

\begin{figure}[t!]
\includegraphics[angle=270,width=1.\columnwidth]{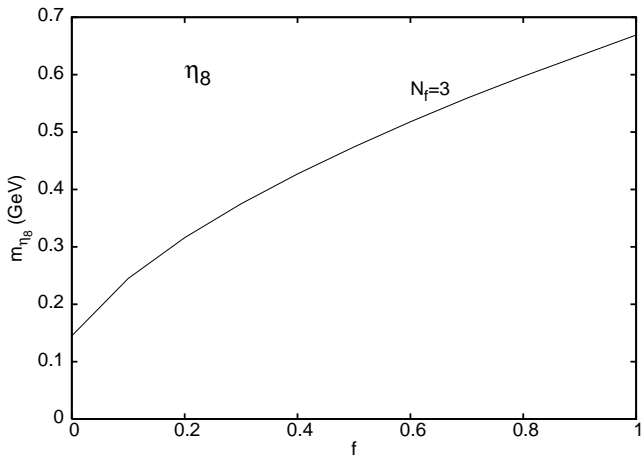}
\caption{Dependence of the $\eta_8$ mass on the "weight" factor {\it
f} for three quark flavors.}\label{eta8}
\end{figure}

The $SU(3)_f$ octet and singlet isospin zero states, $\eta_8$ and
$\eta_0$, can be expressed in the $q\bar{q}$-basis,
\begin{eqnarray}
|\eta_8\rangle&=&\frac{1}{\sqrt{6}}\left(|u\bar{u}\rangle+|d\bar{d}\rangle-2|s\bar{s}\rangle\right),\\
|\eta_0\rangle&=&\frac{1}{\sqrt{3}}\left(|u\bar{u}\rangle+|d\bar{d}\rangle+|s\bar{s}\rangle\right).
\end{eqnarray}
while in $SU(2)_f$ only the $\eta_0$ state is defined.
The "annihilation diagram" does not contribute to the mass of
$\eta_8$ in the chiral limit. For the finite quark masses, however,
it does still contribute due to the mass difference of {\it u}({\it
d}) and {\it s} quarks. For $\eta_0$ this diagram makes a
difference both in chiral limit and in the finite quark mass case.
To find the masses of the $\eta_0$ and $\eta_8$ we have to solve the
system of two coupled integral BSEs (one for {\it u}({\it d}) and
one for {\it s} flavor), equivalent to the equation (\ref{BSE}). An
example of such an equation for the $u\bar{u}$ channel, including its
coupling to the $d\bar{d}$ and $s\bar{s}$ channels, is depicted
diagrammatically in Fig. \ref{uu-bar}.
The dependencies of $\eta_0$ and $\eta_8$ meson masses on the
"weight" factor $f$ are presented in Fig. \ref{eta0} and \ref{eta8}.
The full BSE has been solved for each "weight" factor.

\begin{figure}[b]
\includegraphics[angle=270,width=1.\columnwidth]{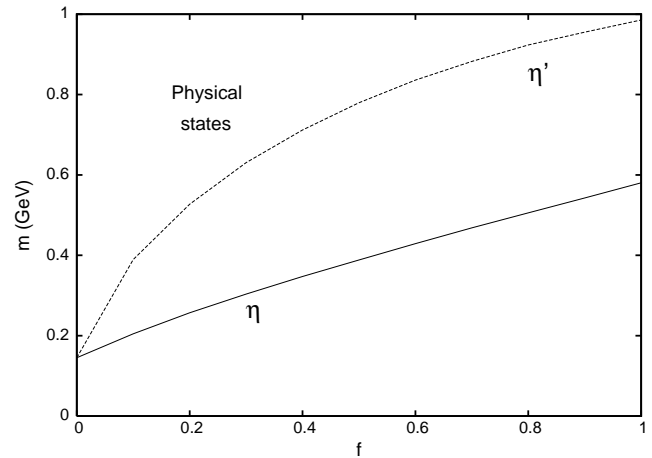}
\caption{Dependence of the physical state masses $\eta$ and $\eta'$
on the "weight" factor {\it f}.}\label{EtaFiniteQuarkMass}
\end{figure}

To make a connection with the physical mass eigenstates, we introduce
the mixing angle $\Theta$,
\begin{eqnarray}
|\eta\rangle&=&\cos{\theta}|\eta_8\rangle-\sin{\theta}|\eta_0\rangle,\\
|\eta'\rangle&=&\sin{\theta}|\eta_8\rangle+\cos{\theta}|\eta_0\rangle \ .
\end{eqnarray}
By comparing the results of our calculation to the experimental
values of $\eta$ and $\eta'$ mass we can determine the mixing angle
$\theta$,
\begin{eqnarray}
\tan^2{\Theta}=\frac{M_{\eta_8}-M_{\eta}}{M_{\eta'}-M_{\eta_8}}\\
\tan^2{\Theta}=\frac{M_{\eta'}-M_{\eta_0}}{M_{\eta_0}-M_{\eta}} \ .
\end{eqnarray}
For the "weight" factor $f=0.9$ we obtain $\Theta\approx -28^0$ in
reasonable agreement with experiment, which favors the mixing angle
in the vicinity of $-20^0$ \cite{Klabucar:1997zi,Yao:2006px}. Now we
can use this angle to explore the dependence of the physical $\eta$
and $\eta'$ on the weight factor. The corresponding results are
presented in Fig.\ref{EtaFiniteQuarkMass}.

\section{Conclusion}\label{conclu}

%

We identify a full one-quark-loop class of microscopic quark diagrams
contributing to the $\eta$ and $\eta'$ masses, separating these
pseudoscalar mesons from the $\pi$ and $K$ mesons.

Because it is not yet possible to compute the full class of
diagrams, we identify the dominant diagrams, equivalent to the
exchange of the ground state pseudoscalar mesons.

Nevertheless this problem still requires state of the art
techniques. We avoid the equal-time framework, where the number of
diagrams would be much larger because the quark propagators would be
separated from the antiquark propagators. We adopt the Euclidian
time Schwinger Dyson framework. The Bethe-Salpeter equation turns
out to be solvable, because the poles in the solution of the gap
equation, for negative Euclidian momentum, does not cross the
complex domain of integration in the Argand plot of the momentum,
depicted in Fig. \ref{IntDom}. Together with the isovector ground
state pseudoscalars \cite{Maris:1997tm,Maris:1997hd,Maris:2000sk},
first excited pseudoscalar states \cite{Holl:2004fr} and ground
state vectors \cite{Maris:1999nt}, this is a new case among the few
where the Euclidian Bethe Salpeter equation has been solved so far.

We notice that the infrared part of the complete class of diagrams
should vanish in the chiral limit. Only the ultraviolet anomaly of
Adler Bell and Jackiw should survive in that limit. Thus it is a
priori expected that our dominant diagram should be excessive to
produce the $\eta$ and $\eta'$ masses, since a cancelation with the
other diagrams should occur in the chiral limit. Hence we multiply
this diagram by a reducing factor $f$. We find that our diagram is
indeed dominant since only the factor $f=0.9$ is necessary to arrive
at the correct experimental masses. Importantly, we then essentially
comply with the experimental mixing of the $\eta_0$ and $\eta_8$.

Our results show that the quark model may solve the U(1) problem
with simple microscopic interactions, although new technical
advances in the computation of full ladders are necessary
before this problem is fully solved. More progresses may
be achieved with the computation of contributions to the
diagrams in Fig.
\ref{PossibleLadderDressingKernel}
(b) and with the separation of the infrared
and of the ultraviolet contributions to the  $\eta$ and $\eta'$
masses.

Importantly we also study the case where there are only two light
flavors. This case is not realistic, nevertheless it is important
to be compared with the two-flavor studied in Lattice QCD. In this
two-flavor case we find that the $\eta'$ mass is only reduced by 20\% .
This stresses the relevance of two-flavor studies in Lattice QCD.

On the other hand, the importance of the one-pion exchange
diagram explains why Lattice QCD, with the present size
of the lattices, cannot directly reproduce the $\eta$ and $\eta'$
masses. Since the pion is very light, a large lattice will
be necessary to encompass with no chiral extrapolation the extent of
the one pion exchange potential between the quarks.

\acknowledgments

PB thanks discussions on the relevance of glueballs with Felipe
Llanes-Estrada. OL would like to thank Peter Tandy, Nicholas
Souchlas, Reinhard Alkofer and Andreas Krassnigg for useful
conversations. PB was supported by the FCT grants POCI/FP/63437/2005
and POCI/FP/63405/2005. OL was supported in part by the U.S.
National Science Foundation under Grant Nos. PHY-0610129.



\begin{thebibliography}{00}



\bibitem{Weinberg:1975ui}
  S.~Weinberg,
  Phys.\ Rev.\  D {\bf 11}, 3583 (1975).


\bibitem{Nambu:1961fr}
  Y.~Nambu and G.~Jona-Lasinio,
  Phys.\ Rev.\  {\bf 124} (1961) 246.

\bibitem{GellMann:1960np}
  M.~Gell-Mann and M.~Levy,
  Nuovo Cim.\  {\bf 16}, 705 (1960).

\bibitem{'t Hooft:1976fv}
  G.~'t Hooft,
  Phys.\ Rev.\  D {\bf 14}, 3432 (1976)
  [Erratum-ibid.\  D {\bf 18}, 2199 (1978)].

\bibitem{Osipov:2005tq}
  A.~A.~Osipov, B.~Hiller and J.~da Providencia,
  Phys.\ Lett.\  B {\bf 634}, 48 (2006)
  [arXiv:hep-ph/0508058].




\bibitem{Duncan:1996ma}
  A.~Duncan, E.~Eichten, S.~Perrucci and H.~Thacker,
  Nucl.\ Phys.\ Proc.\ Suppl.\  {\bf 53}, 256 (1997)
  [arXiv:hep-lat/9608110].

\bibitem{Kuramashi:1994aj}
  Y.~Kuramashi, M.~Fukugita, H.~Mino, M.~Okawa and A.~Ukawa,
  Phys.\ Rev.\ Lett.\  {\bf 72}, 3448 (1994).

\bibitem{Witten:1979vv}
  E.~Witten,
  Nucl.\ Phys.\  B {\bf 156}, 269 (1979).

\bibitem{Aoki:2006xk}
  S.~Aoki {\it et al.}  [JLQCD Collaborations],
  PoS {\bf LAT2006}, 204 (2006)
  [arXiv:hep-lat/0610021].


\bibitem{Bicudo:1998mc}
  P.~J.~A.~Bicudo,
  Phys.\ Rev.\  C {\bf 60}, 035209 (1999)
  [arXiv:nucl-th/9802058].

\bibitem{Fischer:2007ze}
  C.~S.~Fischer, D.~Nickel and J.~Wambach,
  arXiv:0705.4407 [hep-ph].



\bibitem{Adler:1969gk}
  S.~L.~Adler,
  Phys.\ Rev.\  {\bf 177}, 2426 (1969).

\bibitem{Adler:1969er}
  S.~L.~Adler and W.~A.~Bardeen,
  Phys.\ Rev.\  {\bf 182}, 1517 (1969).

\bibitem{Bell:1969ts}
  J.~S.~Bell and R.~Jackiw,
  Nuovo Cim.\  A {\bf 60}, 47 (1969).




\bibitem{'t Hooft:1976up}
  G.~'t Hooft,
  Phys.\ Rev.\ Lett.\  {\bf 37}, 8 (1976).


\bibitem{von Smekal:1997dq}
  L.~von Smekal, A.~Mecke and R.~Alkofer,
  arXiv:hep-ph/9707210.



\bibitem{von Smekal:1997is}
  L.~von Smekal, R.~Alkofer and A.~Hauck,
  Phys.\ Rev.\ Lett.\  {\bf 79}, 3591 (1997)
  [arXiv:hep-ph/9705242].

\bibitem{Maris:2003vk}
  P.~Maris and C.~D.~Roberts,
  Int.\ J.\ Mod.\ Phys.\  E {\bf 12}, 297 (2003)
  [arXiv:nucl-th/0301049].

\bibitem{Holl:2006ni}
  A.~Holl, C.~D.~Roberts and S.~V.~Wright,
  arXiv:nucl-th/0601071.


\bibitem{Bicudo:2003fp}
  P.~Bicudo,
  Phys.\ Rev.\  C {\bf 67}, 035201 (2003)
  [arXiv:hep-ph/0311277].

\bibitem{Llanes-Estrada:2003ha}
  F.~J.~Llanes-Estrada and P.~De A. Bicudo,
  Phys.\ Rev.\  D {\bf 68}, 094014 (2003)
  [arXiv:hep-ph/0306146].

\bibitem{Maris:1999nt}
  P.~Maris and P.~C.~Tandy,
  Phys.\ Rev.\  C {\bf 60}, 055214 (1999)
  [arXiv:nucl-th/9905056].

\bibitem{Maris:1997tm}
  P.~Maris and C.~D.~Roberts,
  Phys.\ Rev.\  C {\bf 56}, 3369 (1997)
  [arXiv:nucl-th/9708029].

\bibitem{Maris:1997hd}
  P.~Maris, C.~D.~Roberts and P.~C.~Tandy,
  Phys.\ Lett.\  B {\bf 420}, 267 (1998)
  [arXiv:nucl-th/9707003].

\bibitem{Maris:2000sk}
  P.~Maris and P.~C.~Tandy,
  Phys.\ Rev.\  C {\bf 62}, 055204 (2000)
  [arXiv:nucl-th/0005015].

\bibitem{Holl:2004fr}
  A.~Holl, A.~Krassnigg and C.~D.~Roberts,
  Phys.\ Rev.\  C {\bf 70}, 042203 (2004)
  [arXiv:nucl-th/0406030].


\bibitem{Klabucar:1997zi}
  D.~Klabucar and D.~Kekez,
  Phys.\ Rev.\  D {\bf 58}, 096003 (1998)
  [arXiv:hep-ph/9710206].

\bibitem{Yao:2006px}
  W.~M.~Yao {\it et al.}  [Particle Data Group],
  J.\ Phys.\ G {\bf 33}, 1 (2006).

\end{thebibliography}
\end{document}